\documentclass[11pt]{article}
\usepackage{moriond,epsfig}

\usepackage{graphicx}
\usepackage{amssymb}
\usepackage{amsfonts}
\usepackage{amsmath}
\usepackage{mathrsfs}
\usepackage{calrsfs}
\usepackage{cite}
\usepackage{pslatex}
\usepackage{hyperref} 

\def\GeVcc{GeV/c$^{2}$}

\def\ra{ \rightarrow }

\newcommand{\sqrts}{\sqrt{s}}

\providecommand{\bbbar}{b\overline{b}}

\providecommand{\pp}{$p\,p$}
\providecommand{\gaga}{\gamma\,\gamma}
\providecommand{\gp}{\gamma\,p}

\begin{document}
\title{Higgs and Beyond the Standard Model physics with the FP420 detector at the LHC}
\author{David d'Enterria}
\address{Laboratory for Nuclear Science, MIT, Cambridge, MA 02139-4307, USA\\
ICC-UB \& ICREA, Universitat de Barcelona, 08028 Barcelona, Catalonia}

\maketitle
\vspace{-0.65cm}
\abstracts{
The physics case of the FP420 R\&D project aiming at the installation of proton detectors 
in the LHC tunnel at 420~m from the ATLAS and CMS interaction points, is presented.
The motivations of the measurements accessible with FP420 -- exclusive Higgs 
production ($p\,p \to p\, H\, p$) and photon-induced processes ($p\,p \to p\,\gamma\, p \to p\, X\, p$,
$p\,p \to p\, \gamma\, \gamma \, p \to p \, X \, p$, where $X$ is sensitive to new physics) -- 
are outlined.
}


\section*{Introduction}

\begin{figure}[htb]
\includegraphics[width=0.32\columnwidth,height=9.0cm]{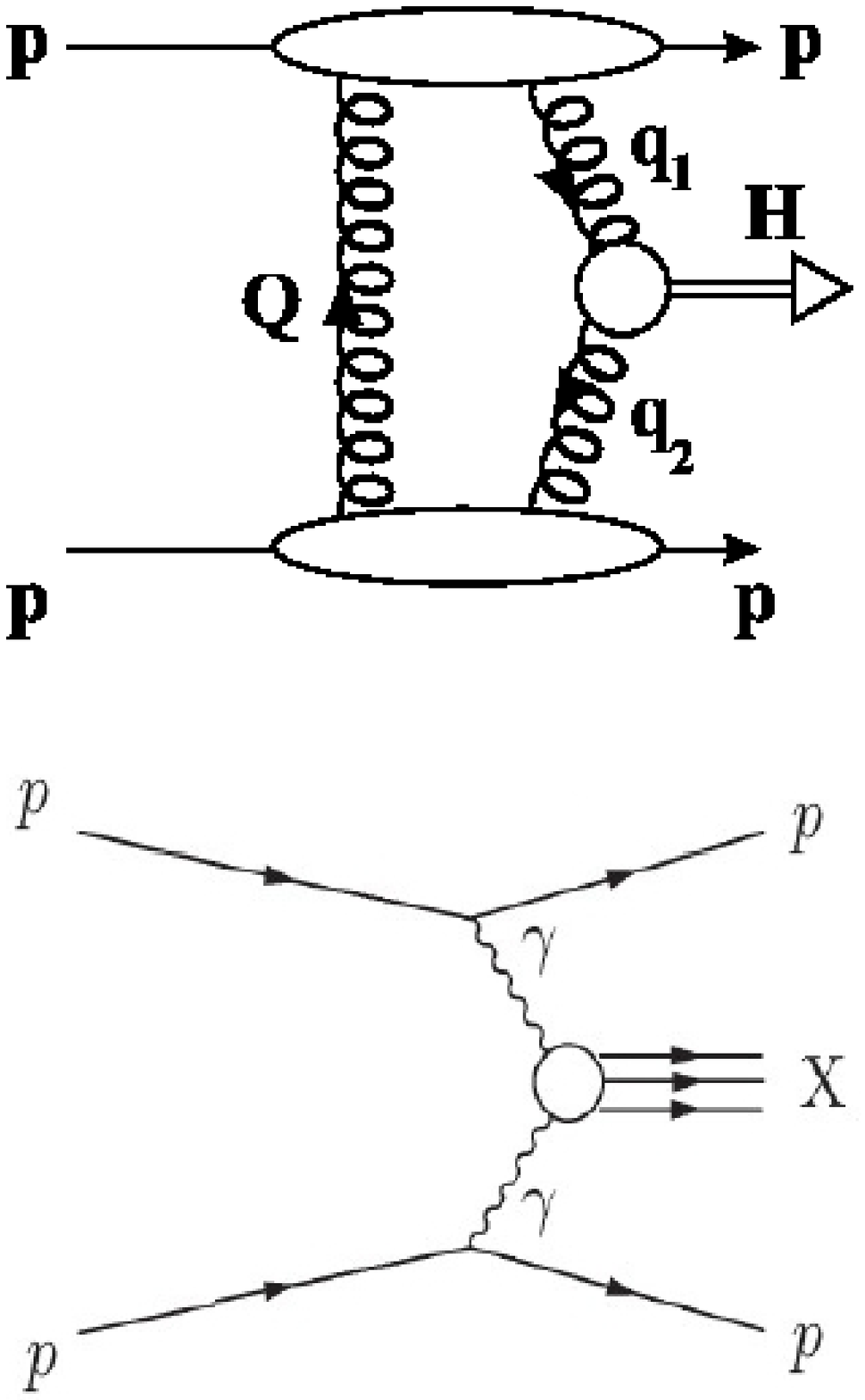}\hspace{0.8cm}
\includegraphics[width=0.62\columnwidth,height=9.8cm]{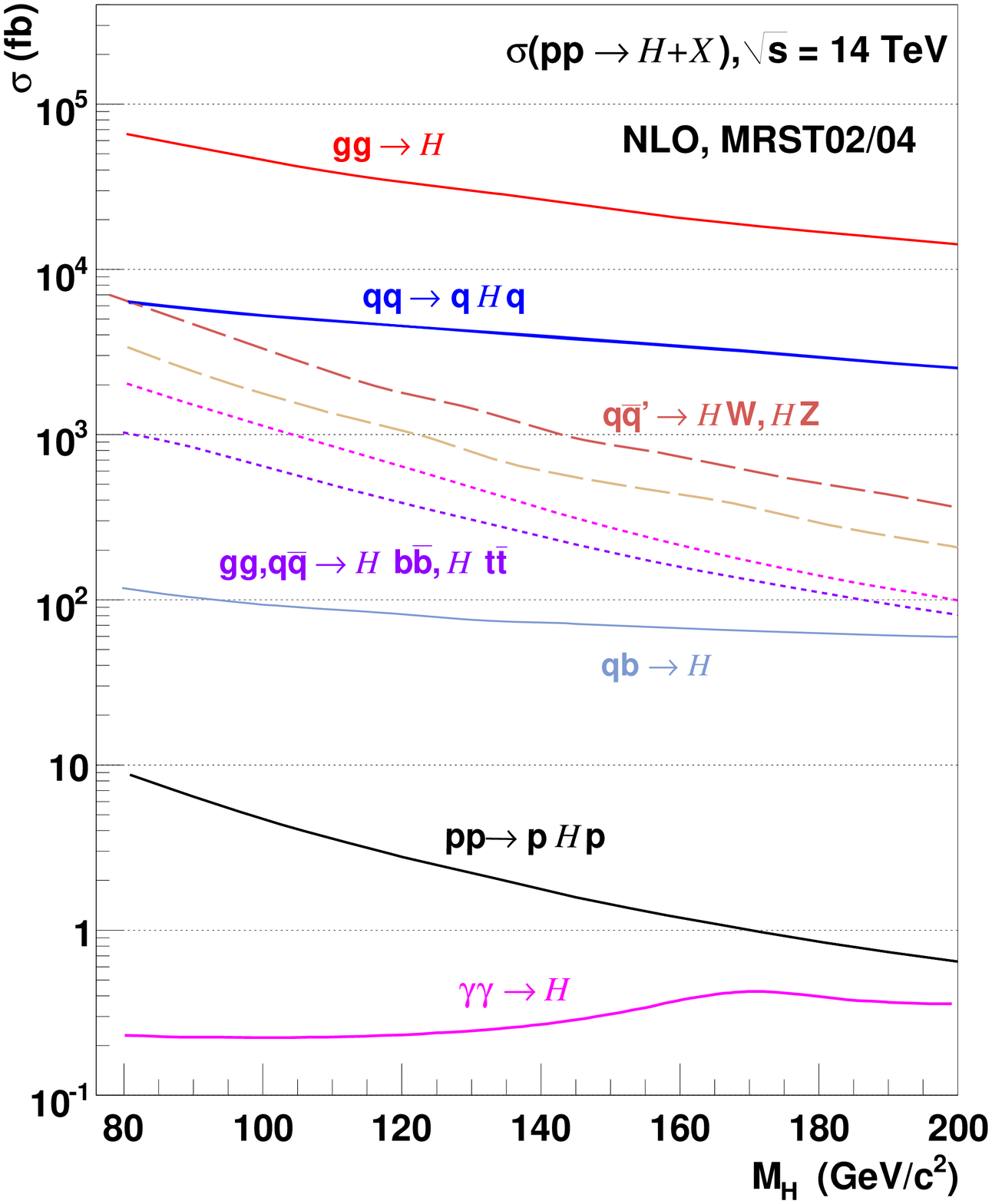}
\caption{Left: Diagrams for central exclusive Higgs production (top) and photon-photon interactions (bottom). 
Right: Cross sections for the SM Higgs versus $M_H$ in production channels accessible at the LHC~\protect\cite{spira}
($\sigma_{pHp}$ is from~\protect\cite{fp420} and $\sigma_{\gaga\to H}$ from~\protect\cite{Papageorgiu:1995eg}).}
\label{fig:1}
\end{figure}

Proton-proton collisions at the LHC will give access to many different scattering processes at energies
never studied before. The primary goal of the collider is the production of new particles predicted within 
or beyond the Standard Model (SM): Higgs boson, SUSY partners of the SM particles ... 
The dominant production mode of heavy particles are high-$p_T$ {\it parton-parton} scatterings in ``head-on'' \pp\ 
collisions. Such collisions are characterised by large QCD activity and backgrounds which often complicate the 
identification of new physics signals. In this context, the clean topologies of {\it exclusive} particle production 
in ``peripheral'' \pp\ processes mediated by colourless exchanges -- such as di-gluon colour-singlet states 
(aka ``Pomerons'') or two photons (Fig.~\ref{fig:1}, left) -- is attracting increasing interest despite their much smaller 
cross sections, $\mathscr{O}(10^{-5})$, compared to the standard parton-parton interactions. Exclusive events are 
characterised by wide rapidity gaps on both sides of the centrally produced system and the survival of both protons 
scattered at very low angles with respect to the beam. The final-state is thus much cleaner, with a larger
signal/background and the event kinematics can be constrained measuring the final protons.\\

\noindent
A prime process of interest is Central Exclusive Production (CEP) of the Higgs boson, $pp \rightarrow p \oplus H \oplus p$, 
('$\oplus$' represents a rapidity-gap i.e. a large region devoid of hadronic activity). In order to detect 
both protons in the range of momentum loss appropriate for Higgs masses close to the LEP limit, 
114~\GeVcc, detectors must be installed a few mm's away to the outgoing beams in the high-dispersion 
region 420~m away from the interaction points on each side of the ATLAS and CMS experiments. 
The FP420 R\&D project\cite{fp420} 
aims at assessing the feasibility of installing such near-beam detectors in the LHC tunnel 
to measure the leading protons 
issuing from central-exclusive or photon-exchange processes, in conjunction with 
the produced system measured in the central ATLAS and CMS detectors.

\section{Higgs physics}


The ATLAS~\cite{atlas_tdr} and CMS~\cite{cms_tdr} experiments have been designed
to discover the SM Higgs, if it exists, in \pp\ at $\sqrts$~=~14~TeV with a few tens of fb$^{-1}$
in at least one or two decay modes, in the dominant gluon-gluon or vector-boson-fusion (VBF) 
production channels (Fig.~\ref{fig:1}, right). As important as discovering a Higgs-like resonance, 
it is to characterise its properties and confirm that it is the particle responsible of the SM 
electroweak symmetry breaking. The following Higgs properties 
are very challenging to determine in the traditional LHC searches:
\begin{itemize}
\item Coupling to $b$-quark: Testing the mass-dependent Yukawa couplings 
of Higgs to the various SM fields is crucial. Yet, the $H\to\bbbar$ decay channel is now 
considered unaccessible at the LHC~\cite{atlas_tdr,cms_tdr} due to the overwhelming 
QCD background: $\sigma(H\to \bbbar)\approx$~20~pb $\ll \sigma(\bbbar)\approx$~500~$\mu$b.
\item Quantum numbers: The expected SM Higgs spin-parity $J^{PC} = 0^{++}$ 
is very difficult to determine at the LHC in the currently favoured range of masses\footnote{At 
higher masses, the azimuthal asymmetry of the SM Higgs VBF-jets or $ZZ$-decay can be used
to confirm its $CP$ numbers.} $M_H\lesssim$~180~\GeVcc.
\item Nearly-degenerate Higgs bosons: Possible additional Higgs states with similar masses
(but different parities) predicted in various extensions of the SM, are not easy to separate.
\item Mass \& width in invisible decays:  Details on possible invisible branching ratios remain
very difficult as their presence can only be determined in counting-type measurements
in VBF channels.
\item (N)MSSM Higgs'es: Additional Higgs bosons in (next-to) minimal supersymmetric 
extensions of the SM, with largely enhanced third-family ($b,\tau$) decays, with 
$CP$-violating mixings, or with complicated decay channels (e.g. $h\to aa \to 4\tau$) remain 
very problematic (if possible at all).
\end{itemize}

The main motivation of a CEP Higgs measurement is that it can help to address all these 
issues well before a possible future $e^+e^-$ linear collider becomes operational. 
Indeed, observation of a Higgs-like resonance in the CEP channel benefits from 
(i) enhanced signal over backgrounds (giving access to the difficult $\bbbar$ decay 
channel)~\cite{DeRoeck:2002hk,Kaidalov:2003ys,cox2,krs2,kmrN};
(ii) quantum numbers measurement via azimuthal asymmetry of the leading protons~\cite{Khoze:2004rc};
(iii) mass determination with  $\mathscr{O}$(2~\GeVcc) resolution from the leading 
protons via the ``missing mass'' method, $M_h^2=(p_1+p_2-p'_1-p'_2)$~\cite{Albrow:2000na}, 
irrespective of the (e.g. possibly invisible~\cite{Belotsky:2004ex}) decay modes; (iv) separation of scalar from pseudoscalar 
degenerate states, as the CEP system is scalar with an approximate $J^{PC} = 0^{++}$ 
selection rule~\cite{Khoze:2000jm}; and (v) discovery possibilities ($\bbbar,\tau\tau$ decays) 
in complicated regions of the MSSM~\cite{Kaidalov:2003ys,Heinemeyer:2007tu,Cox:2007sw} 
or NMSMM~\cite{Forshaw:2007ra}.\\

The CEP Higgs process (top diagram in Fig.~\ref{fig:1}) is dominated by a hard scale, 
$\Lambda_{\rm QCD}^2\ll Q^2\ll M_H^2$, and thus calculable with perturbative QCD
techniques. The QCD factorization theorem is applicable with the addition of an extra factor accounting 
for non-perturbative effects (see below). Schematically,
\[\sigma_{pp\to pHp} = uPDF(x_{1,2},Q^2) \oplus \sigma_{gg\to H} \oplus S^2_{\mbox{\tiny gap survival}}.\]
Here $uPDF(x_{1,2},Q^2)$ are proton 'unintegrated parton distribution functions' which can be approximated
in terms of standard gluon distribution functions, $g(x,Q^2)$ evaluated at $x=M_H/\sqrts$ and $Q=M_H/2$, 
times a 'Sudakov suppression factor' encoding the probability that the fusing gluons do not radiate 
in their evolution from $Q$ up to the hard scale. The possibility of soft rescatterings where particles 
from the underlying \pp\ event (i.e. from other parton interactions) populate the gaps, is basically independent 
of the short-distance subprocess and can be taken into account with a multiplicative gap survival probability 
factor $S^2$, 
computable within eikonal approaches~\cite{KMRsoft,Ryskin:2007qx}. For $S^2$~=~0.03,
the expected SM CEP Higgs cross section ($M_H$~=~120~\GeVcc) is around 3~fb (Fig.~\ref{fig:1}, right). 
The reliability of such theoretical calculations has been cross-checked at the Tevatron in the 
exclusive production of high-mass dijets, $p\bar{p} \ra p + jj + \bar{p}$~\cite{Aaltonen:2007hs} 
and scalar quarkonium states $p\bar{p} \ra p +\chi_{c0}  + \bar{p}$~\cite{Aaltonen:2009kg},
which are well described by the theory. In certain regions of the MSSM, at high $\tan\beta$ and 
small $M_A$, with enhanced Higgs coupling to fermions, $\sigma_{pp\to pHp}$ can be a factor 
of 10--100 larger~\cite{Kaidalov:2003ys,Heinemeyer:2007tu}.

\begin{figure}[htbp]
\includegraphics[width=0.60\columnwidth,height=5.0cm]{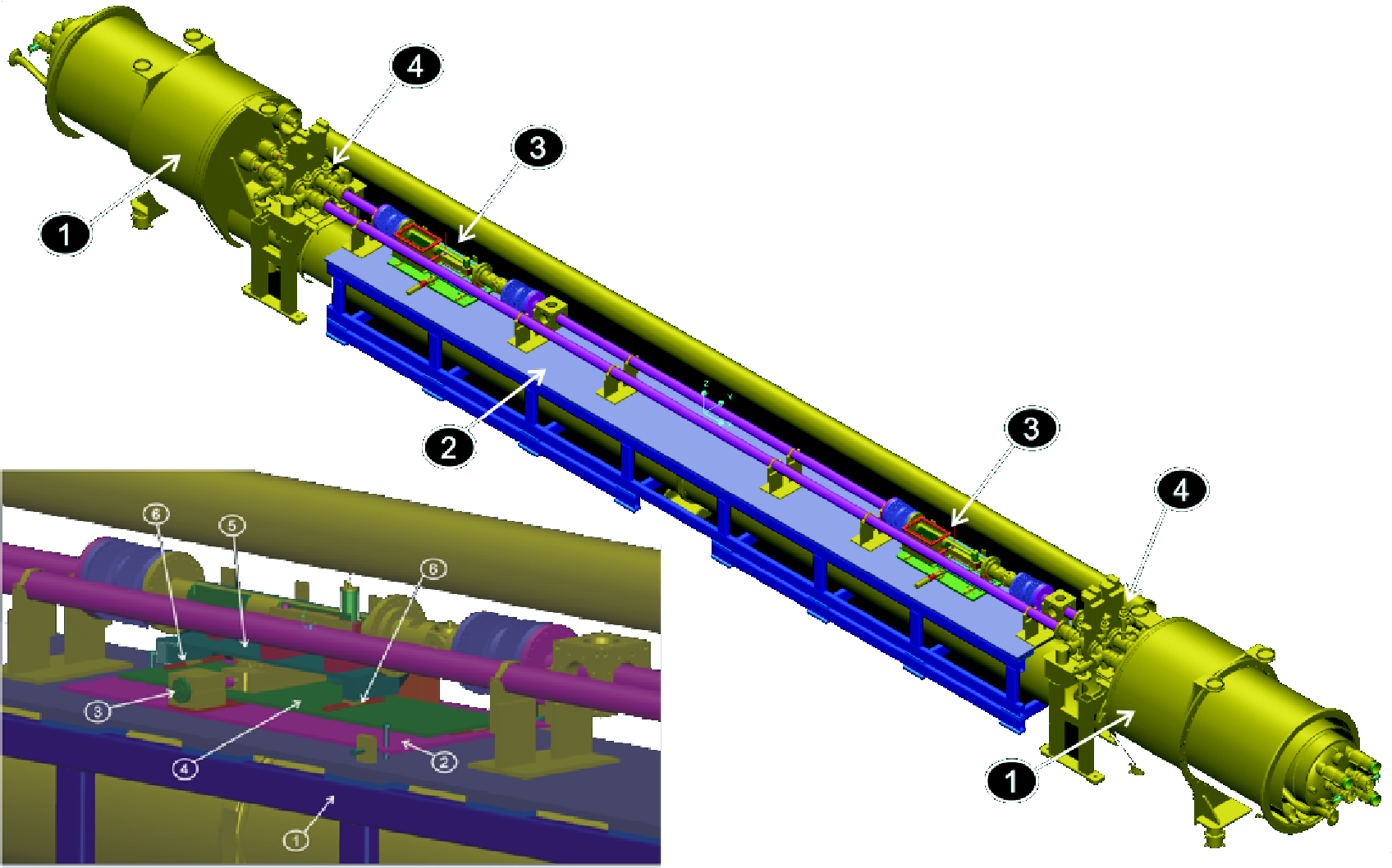}
\includegraphics[width=0.40\columnwidth,height=5.0cm]{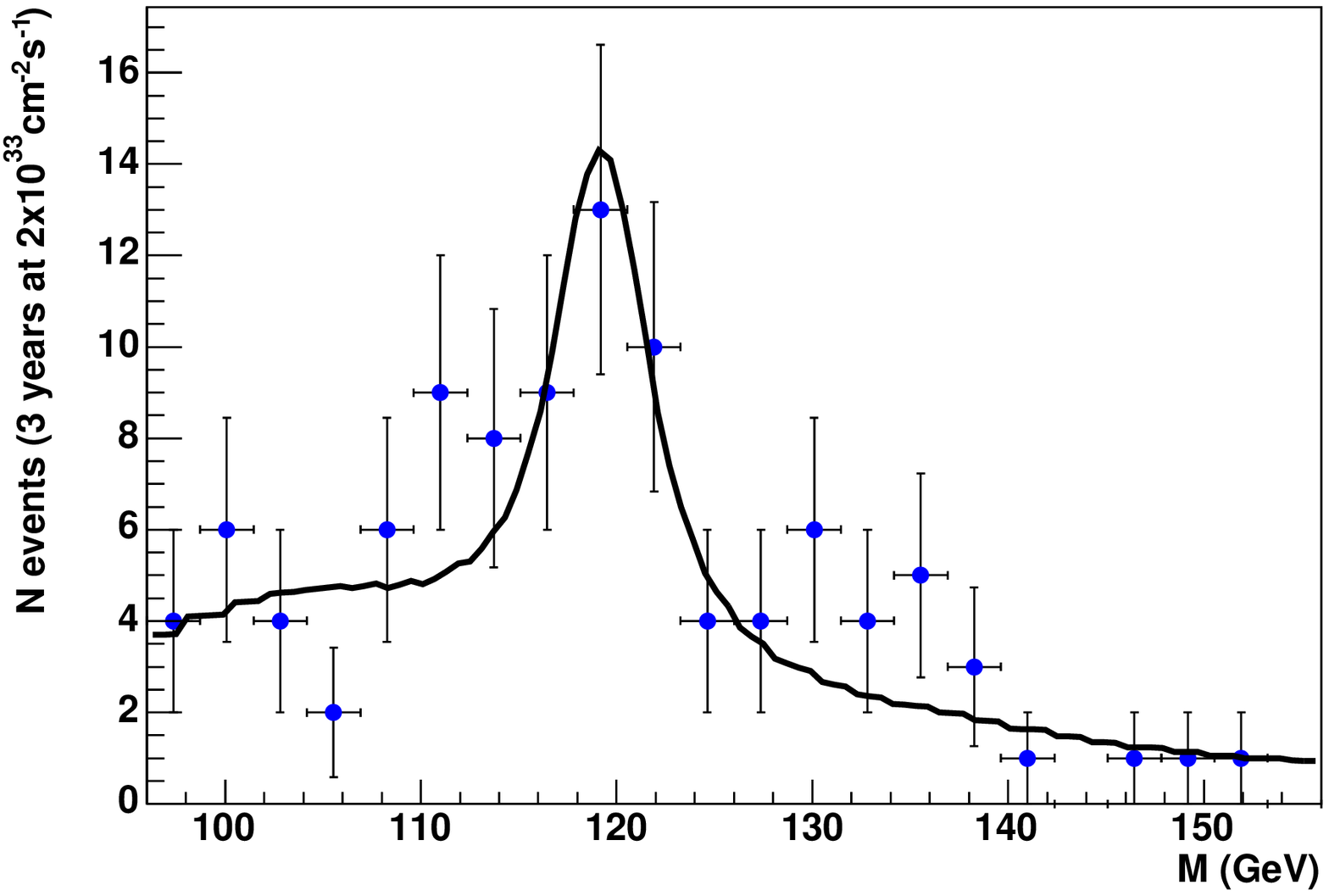}
\caption{Left: Top view of the proposed FP420 system on top of the moving 'Hamburg' pipe in the cold
area of the LHC tunnel (zoom shows the support table with one detector section). 
Right: Expected mass fit for the MSSM $h\rightarrow b\bar{b}$ decay ($M_H$~=~120~\GeVcc), 
measured with FP420 in 60~fb$^{-1}$ integrated luminosity~\protect\cite{Cox:2007sw}.
(The significance of the fit is $3.5 \sigma$).}
\label{fig:fp420_CEP}
\end{figure}

For Higgs masses close to the LEP limit, $M_H\approx$~120~\GeVcc, both protons lose a longitudinal 
momentum fraction $\xi_{1,2}\approx$~1\% (using $M_X^2\approx\xi_1\xi_2 \,s$)
and, after accounting for the LHC beam optics~\cite{hector}, the optimal proton tagging acceptance 
is beyond the current ALFA~\cite{alfa} and TOTEM~\cite{loi_cms_totem} Roman Pots (RPs) detectors 
around 220~m (which have $p$ acceptances for larger masses, $0.02<\xi<0.2$).
The proposed FP420 detector system~\cite{fp420} -- a magnetic spectrometer consisting of a moveable 
3-D silicon tracking system and fast \v{C}erenkov detectors located in a 12-m-long region at about 420~m from 
the ATLAS and CMS IPs (Fig.~\ref{fig:fp420_CEP}, left) --  allows for the detection of both outgoing protons 
scattered by a few hundreds $\mu$rads (i.e. 3 -- 9~mm at 420~m) relative to the LHC beamline.
A measurement of the protons relative-time of arrival in the 10~ps range is required 
for matching them with a central vertex within $\sim$2 mm. Such a vertex matching is required
to reject a large fraction of the simultaneous \pp\ pile-up collisions at high-luminosities. Under such
circumstances, MSSM Higgs line-shapes can be reconstructed e.g. in the $b\bar{b}$ channel with a 
$3\sigma$ or better significance with an integrated luminosity of 60~fb$^{-1}$
(Fig.~\ref{fig:fp420_CEP}, right)~\cite{Cox:2007sw}.


\section{Photon-proton and photon-photon physics}

\begin{figure}[htbp]
\includegraphics[width=0.50\columnwidth,height=7.5cm]{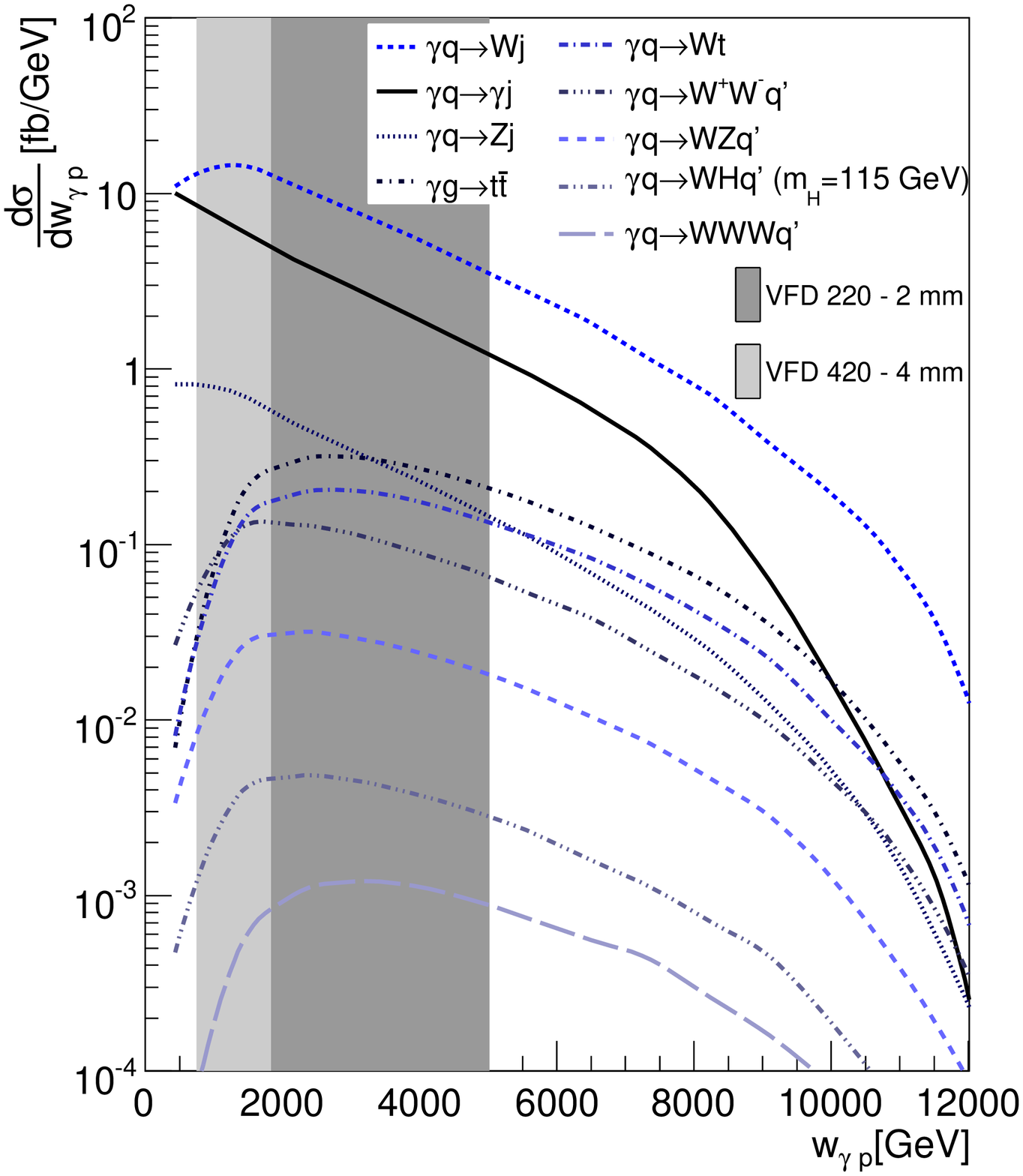}\hspace{0.5cm}
\includegraphics[width=0.50\columnwidth,height=7.5cm]{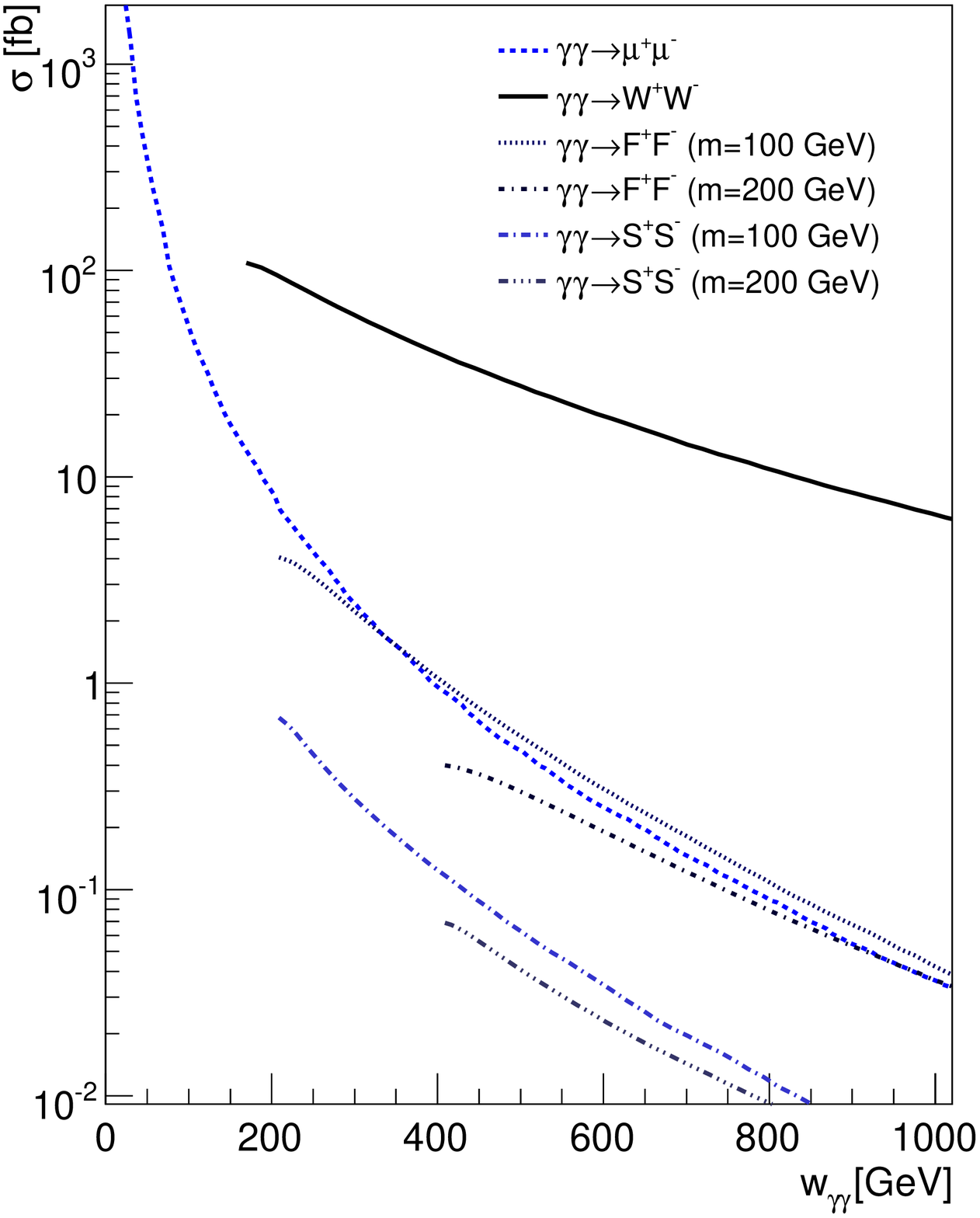}
\caption{Left: Differential cross-sections for various $pp (\gamma q/g \rightarrow X) p Y$ processes as 
a function of the c.m. energy in $\gamma$-proton collisions, $W_{\gamma p}$. The acceptance of RPs
(220~m at 2~mm from the beam-axis, and 420~m at 4~mm from the beam-axis) is also sketched~\cite{louvain}.
Right: Cross sections for various $\gamma \gamma$ processes  at the LHC as a function of the minimal 
$W_{\gaga}$ c.m. energy~\cite{louvain}.}
\label{fig:gammap_gammagamma}
\end{figure}

A significant fraction of $pp$ collisions at the LHC will also involve quasi-real (low-$Q^2$) 
photon interactions: one photon is emitted by one (or both) incoming proton(s) which then
subsequently collides with the other proton (photon) producing a system $X$ (Fig.~\ref{fig:1}, bottom left).
The LHC thus offers the unique possibility to study $\gaga$ and $\gp$ processes 
at centre-of-mass (c.m.) energies well beyond the electroweak scale. 
Such photon-induced interactions are less central (i.e. take place at larger impact parameters)
than Pomeron-induced processes and, thus, the exchanged squared-momentum $t\approx p_T^2$ 
is smaller\footnote{The $t$-distribution is of the type $\exp(-b\;t)$ with slope $b\approx$~4 (40) GeV$^{-2}$
for double-Pomeron (double-photon) collisions.}.
Differential cross sections for $pp (\gamma q/g \rightarrow X) p Y$ reactions, as a function 
of the $\gamma$-proton c.m. energy, are presented in Fig.~\ref{fig:gammap_gammagamma} 
(left) together with the acceptance of forward proton taggers. A large variety of processes have sizeable 
cross section well in the TeV energy range.  Fig.~\ref{fig:gammap_gammagamma} (right) shows
various pair production cross sections (for charged and colourless fermions and scalars of two 
different masses) as a function of the minimal photon-photon c.m. energy $W_{\gaga}$.

Various exclusive photon-induced processes sensitive to new physics 
are accessible to measurement at the LHC with forward proton taggers~\cite{louvain}:
\begin{enumerate}
\item two-photon production of $W$ and $Z$ pairs sensitive to anomalous quartic gauge couplings~\cite{Pierzchala:2008xc},\vspace{-0.2cm}
\item two-photon production of supersymmetric pairs~\cite{Schul:2008sr}, and\vspace{-0.2cm} 
\item anomalous single top photoproduction~\cite{deFavereau:2008hf}.
\end{enumerate}


Many physics scenarios beyond the SM, with novel interactions and/or particles, lead to 
modifications of the gauge boson ($\gamma$, $W$ and $Z$) self-interaction vertices. 
Two-photon production of $W$,$Z$ pairs provides an excellent opportunity to investigate 
anomalous {\it quartic} gauge couplings. The $WW$ process has a 
total cross section of more than 100~fb, and a very clear experimental signature. 
The processes  $\gamma\gamma\rightarrow W^+W^-\rightarrow l^+l^-\nu\bar{\nu}$ and 
$\gamma\gamma\rightarrow ZZ\rightarrow l^+l^- j j$ have been investigated via the 
signature of two leptons ($e$ or $\mu$) within the CMS~\cite{Pierzchala:2008xc} 
acceptance. The calculated cross section upper limits can 
then be converted to limits on the anomalous quartic couplings which are about 4000 times 
stronger than the best limits established at LEP2.


The 
SUSY pair cross-section at the LHC, e.g. $\gamma \gamma \rightarrow \tilde{l}^{+} \tilde{l}^{-}$, 
has 
cross-sections $\mathcal{O}(20$~fb$)$
still consistent with the LEP search limits. 
Two-photon exclusive production of pairs of new charged particles benefits from 
(i) the possibility to significantly constrain their masses, using double leading-proton information, and
(ii) in the case of SUSY pairs, the presence of simple final states without cascade decays, characterised by 
two (acoplanar) charged leptons with large missing energy 
with low backgrounds, and large trigger efficiencies. With this technique and sufficient statistics, 
masses could be measured with precision of a few \GeVcc~by looking at the 
minimal c.m. energy required to produce the heavy pair~\cite{Schul:2008sr}. 

Single top photo-production in the SM is only possible for higher-order electroweak 
interactions, since neutral currents preserve quarks flavour at tree level. The observation of a 
large number of single top events would hence be a sign of Flavour Changing Neutral Currents 
(FCNC) induced by processes beyond the SM.  A general effective Lagrangian for such processes 
can be written with anomalous couplings $k_{tu\gamma}$ and $k_{tc\gamma}$.
Strong limits on the anomalous coupling $k_{tu\gamma}$ (of which the current best value, 
from ZEUS, is around 0.14) and the unprobed $k_{tc\gamma}$ can be obtained with
forward proton taggers in photon-proton collisions after 1~fb$^{-1}$ of integrated 
luminosity~\cite{deFavereau:2008hf}.


\section*{Acknowledgments}

I would like to thank the organizers of Moriond'09 for their kind invitation to this unique meeting. 
I am grateful to Albert de Roeck, Valery Khoze and Andy Pilkington for informative discussions. 


\end{document}